\documentclass[11pt,reqno]{amsart}
\usepackage{amsfonts,amsmath,amssymb}
\newtheorem{Theorem}{Theorem}

\def\beq#1#2\eeq{%
        \begin{equation}%
        \label{#1}%
            #2%
        \end{equation}%
    }

\usepackage{color}
\usepackage{graphicx}
\usepackage{epstopdf}

\theoremstyle{plain}
\newtheorem{theorem}{Theorem}

\theoremstyle{remark}
\newtheorem{remark}[theorem]{Remark}

\theoremstyle{definition}

\def\Im{\mathop{\rm Im}}

\title[Euler-Dirac)]{Integrable generalisations of Dirac magnetic monopole}

\author{A.P. Veselov}
\address{Department of Mathematical Sciences,
Loughborough University, Loughborough LE11 3TU, UK;  Moscow State University and Steklov Mathematical Institute, Moscow, Russia}
\email{A.P.Veselov@lboro.ac.uk}

\author{Y. Ye}
\address{Department of Mathematical Sciences,
Loughborough University, Loughborough LE11 3TU, UK}
\email{Y.Ye@lboro.ac.uk}

\begin{document}

\maketitle

\begin{abstract}
We classify certain integrable (both classical and quantum) generalisations of Dirac magnetic monopole on topological sphere $S^2$ with constant magnetic field, completing the previous local results by Ferapontov, Sayles and Veselov.

We show that there are two integrable families of such generalisations with integrals, which are quadratic in momenta. The first family corresponds to the classical Clebsch systems, which can be interpreted as Dirac magnetic monopole in harmonic electric field. The second family is new and can be written in terms of elliptic functions on sphere $S^2$ with very special metrics.

\end{abstract}

\tableofcontents

\section{Introduction}

The history of quantum integrable systems with magnetic fields goes back to the pioneering work in the 1930s by Dirac \cite{Dirac} on the celebrated magnetic monopole and by Landau \cite{Landau}, who considered the case of constant magnetic field on the plane (Landau problem).  Since then this area was of a substantial interest of the mathematical and theoretical physicists (see e.g. \cite{Fer-Ves,KV,Mlad,NP,Prieto,Sol,WY}).

In spite of this, the general problem of quantum integrability in two dimensions in the presence of a
magnetic field is still far from complete solution.  Some important results in this direction have been found, in particular, by Winternitz and his collaborators in
\cite{ber-wint, DGRW,MW}.

Ferapontov and Fordy \cite{FF} derived the classical integrability conditions in the case, when the integral is quadratic in momenta.
Ferapontov, Sayles and Veselov \cite{FSV} considered the quantum case and showed that the conditions of quantum integrability are different from classical case (see the details in the next section).
However, remarkably they coincide in the case when the density of the magnetic field $B$ is constant. This case was studied in \cite{FSV}, where a local classification of such systems  under some additional assumptions was found. 

The final list consists of two families (see the next section). The first one contains the Dirac magnetic monopoles in the external harmonic field (and their hyperbolic versions), which are known to be equivalent to the classical Clebsch integrable cases of the free rigid body in infinite ideal fluid \cite{Clebsch}.

The second family is more mysterious and is the main object of our study. We show that under certain assumptions on the parameters the corresponding systems can be extended to the smooth systems on the topological sphere $\mathbb S^2$, which can be described in terms of elliptic functions.

More precisely, we represent the sphere $\mathbb S^2$ as the quotient of a real torus $\mathbb T$ by the involution $\sigma: u\to -u,\, u \in \mathbb T.$

Consider the elliptic function $\mathcal Q(z)$ defined as the inversion $w=\mathcal Q(z)$ of the elliptic integral
$$
z=\int_{\beta_2}^{w}\frac{2d\xi}{\sqrt{P(\xi)}},
$$
where 
$$
P(x)=a_3(x-\beta_1)(x-\beta_2)(x-\beta_3)(x-\beta_4)
$$
is a polynomial with $a_3<0$ and 4 real roots: $\beta_1>\beta_2>0>\beta_3>\beta_4,$
such that
$$\beta_1+\beta_2+\beta_3+\beta_4=0, \, \beta_1+\beta_4<0,\,\, \beta_2+\beta_3>0.$$
The elliptic function $\mathcal Q(z)$ is even and 
has two periods: real $2K_1$ and pure imaginary $2iK_2$, where
$$
K_1=\int_{\beta_2}^{\beta_1}\frac{2d\xi}{\sqrt{P(\xi)}}, \,\,\, K_2=\int_{\beta_2}^{\beta_3}\frac{2d\xi}{\sqrt{-P(\xi)}}.
$$
It satisfies the differential equation
$
4\mathcal Q'^2=P(\mathcal Q)
$
and can be expressed via the standard Weierstrass elliptic function $\wp(z)$.
In the limiting case when $\beta_1+\beta_4=\beta_2+\beta_3=0$ (so $P(x)$ is even), $\mathcal Q$ can be written in terms of the Jacobi's elliptic $sn$-function \cite{WW} as
$$
\mathcal Q=\beta_2 \, \textit{sn}(\alpha (z-\beta_2);k),\,\, \alpha=\sqrt{a_3}\beta_1/2, \,\, k=\beta_2/\beta_1.
$$
Introduce two real-valued functions
$$Q_1(u_1):=\mathcal Q(u_1),\,\,\,\,\, Q_2(u_2):=\mathcal Q(iu_2)$$
with periods $2K_1$ and $2K_2$ respectively, and consider the torus 
$$\mathbb T^2=\mathbb R^2(u_1,u_2)/4K_1 \mathbb Z\oplus 4K_2 \mathbb Z.$$

On this torus the corresponding classical Hamiltonian $H$ and integral $F$ have the following explicit form with $Q_1=Q_1(u_1)$ and $Q_2=Q_2(u_2)$
\begin{equation}
\label{H1}
H=\frac{1}{Q_1^2-Q_2^2}\left[(p_1-A_1)^2+(p_2-A_2)^2\right]+\frac{\mu}{Q_1+Q_2},
\end{equation}
\begin{equation}
\label{F1}
F=\frac{1}{Q_1^2-Q_2^2}\left[Q_2^2(p_1-A_1)^2+Q_1^2(p_2-A_2)^2\right]
\end{equation}
$$
+\frac{2kQ_2'}{Q_1-Q_2}(p_1-A_1)+\frac{2kQ_1'}{Q_2-Q_1}(p_2-A_2)-\frac{\mu Q_1Q_2}{Q_1+Q_2}-kB(Q_1+Q_2)^2,
$$
where $B$ is the density of magnetic field assumed to be constant and $k=-4B/a_3.$ The magnetic potential $A=A_1du_1+A_2du_2$ is determined by the relation
\begin{equation}
\label{A1}
dA=B(Q_1^2(u_1)-Q_2^2(u_2))du_1\wedge du_2.
\end{equation}
There is a problem with these systems on the torus, because $Q_1-Q_2=0$ at the half-periods of the torus, which creates singularities in the formulas.
However, we show that on the quotient $\mathbb S^2=\mathbb T^2/\sigma$ of the torus by involution $\sigma$ having exactly these points fixed, this problem disappears and we have regular smooth systems on $\mathbb S^2.$

In the limiting even case we do have two singularities in the potential $h$, but the metric becomes the standard metric on the round sphere, so we have the new integrable electric perturbation of Dirac magnetic monopole (and new integrable two-centre problem) on the standard sphere (see \cite{VY}).

The plan of the paper is following. In the next two sections we describe the classical and quantum integrability conditions in 2D in the presence of magnetic field and prove the local classification result in the case of non-zero constant magnetic field, mainly following unpublished work of Ferapontov, Sayles and Veselov \cite{FSV}. Then we show that under certain condition on the parameters these systems can be extended to the regular analytic integrable systems on the topological sphere $S^2$ with some very special metrics. 

\section{Integrable magnetic fields in 2D: local classification}

In two dimensions it is always possible to reduce both Hamiltonian $H$ and integral $F$ to a diagonal form:
\begin{eqnarray}
   H&=&g^{11} \left(p_1-A_1\right)^2+ g^{22}
\left(p_2-A_2\right)^2+h,\nonumber\\[-4mm]
\label{classic_hams}\\[-2mm]
   F&=& g^{11} v^1\left(p_1-A_1\right)^2+ g^{22}v^2
\left(p_2-A_2\right)^2 +\phi^1 \left(p_1-A_1\right)
+\phi^2\left(p_2-A_2\right) + \varphi, \nonumber
\end{eqnarray} 
in which metric $g^{ii}$ and all the other coefficients $v^i$, $A_i$, $\phi^i$, $h$, $\varphi$ are functions depending on the coordinates $(q^1,q^2)$. 

Ferapontov and Fordy \cite{FF} showed that Poisson
commutativity of $H$ and $F$ is equivalent to the following integrability conditions
\begin{eqnarray}
&& (C1) \ \ \ \partial_i v^i = 0, \ \ \ \ \,\, i=1,2, \nonumber\\[3mm]
&& (C2) \ \ \ \partial_j v^i= \left(v^j-v^i\right)  \partial_j \ln(g^{ii}) \ \ \ \
\mbox{for all} \ i \neq j, \nonumber\\[3mm]
&& (C3) \ \ \ \partial_i \phi^i = \frac{1}{2 g^{ii}} \left( \phi^1 \partial_1
g^{ii} + \phi^2 \partial_2 g^{ii}\right), \,\, \ i=1,2,  \qquad\qquad\qquad \qquad\qquad\quad \ \; \nonumber\\[4mm]
&& (C4) \ \ \ 2 \sqrt{g^{11} g^{22}}(v^2-v^1) B = g^{22} \partial_2
\phi^1 +g^{11} \partial_1 \phi^2 , \label{full_prob_with_B}\\[4mm]
&& (C5) \ \ \ \partial_1 \varphi - v^1 \partial_1 h - \frac{\phi^2}{\sqrt{g^{11} g^{22}}} B =0, \quad \partial_2 \varphi- v^2 \partial_2  h + \frac{\phi^1}{\sqrt{g^{11} g^{22}}} B =0, \nonumber\\[2mm]
&& (C6) \ \ \ \phi^1 \partial_1 h + \phi^2 \partial_2 h=0, \nonumber
\end{eqnarray}
where
\begin{equation}
\label{mag}
B:=\sqrt{g^{11} g^{22}}(\partial_1 A_2 -\partial_2 A_1) 
\end{equation}
is the magnetic field density.

Consider now the following quantum analogue of the Hamiltonian and the integral:
\begin{equation}
\label{q_ops}
   \hat{H} = \sqrt{g^{11}g^{22}}\nabla_1\frac{g^{11}}{\sqrt{g^{11}g^{22}}}\nabla_1
  +  \sqrt{g^{11}g^{22}}\nabla_2\frac{g^{22}}{\sqrt{g^{11}g^{22}}}\nabla_2 + h,
 \end{equation}
 $$
 \hat{F}= v^1\sqrt{g^{11}g^{22}}\nabla_1\frac{g^{11}}{\sqrt{g^{11}g^{22}}}\nabla_1
  + v^2\sqrt{g^{11}g^{22}}\nabla_2\frac{g^{22}}{\sqrt{g^{11}g^{22}}}\nabla_2
 +\phi^1 \nabla_1 +\phi^2\nabla_2 + \varphi, 
$$
where $\nabla_j=i\partial_j-A_j, \, j=1,2$.

Ferapontov, Sayles and Veselov \cite{FSV} derived the necessary and sufficient conditions for commutativity $[\hat H, \hat F]=0$ and showed that the first conditions (C1)-(C5) are the same, but the last condition (C6) in quantum case is replaced by
\begin{equation}
\label{qcond}
(C6)^*\ \ \ \phi^1 \partial_1 h + \phi^2 \partial_2 h+ \sqrt{g^{11} g^{22}} \left(v^2-v^1\right)\left( \frac{\partial_2 g^{11}}{g^{11}} \partial_1 B +\frac{\partial_1 g^{22}}{g^{22}}\partial_2 B-\partial_1 \partial_2 B \right)  =0.
\end{equation}

In particular, we see that if the magnetic density $B$ is constant then the extra term
$$
 \frac{\partial_2 g^{11}}{g^{11}} \partial_1 B +\frac{\partial_1 g^{22}}{g^{22}}\partial_2 B-\partial_1 \partial_2 B =0
$$
vanishes and the quantum and classical integrability conditions coincide. 

Local classification of all such systems (under some additional assumptions) was done by Ferapontov, Sayles and Veselov \cite{FSV}, who proved in the quantum case the following

\begin{Theorem}
Suppose that the quantum system with the Hamiltonian $\hat{H}$ of the form (\ref{q_ops}) has 
magnetic field with a constant non-zero density $B$, a non-constant electric potential $h$ and assume that the system has no integrals, which are linear in momenta.
	
Then the system has a second order integral $\hat{F}$ if and only if it can be locally reduced to one of the forms specified below, where in each case the metric is of St\"ackel form 
\begin{equation}
\label{metric}
ds^2=\frac{q^1-q^2}{f(q^1)}(dq^1)^2+\frac{q^2-q^1}{f(q^2)}(dq^2)^2
\end{equation}
 with
\begin{eqnarray}
\label{I}
\mbox{\textit{\textrm{(I)}}} && f(q)=a_3 q^3+a_2 q^2+a_1 q+a_0,\,\,\,\, h=\mu(q^1+q^2);\\
\label{II}
\mbox{\textit{\textrm{(II)}}} && f(q)=a_3 q^3+a_2 q^2+a_1q+a_0 q^{\frac32},\,\,\,  h=\frac{\mu}{\sqrt{q^1}+\sqrt{q^2}}
\end{eqnarray}
depending on real parameters $a_0, a_1, a_2, a_3\neq 0$ and $\mu.$ The Gaussian curvature of the metrics respectively is
\begin{equation}
\label{GK}
{\textit{\textrm{(I)}}} \,\,\, K=-\frac{a_3}{4} \quad {\textrm{and \,\,\, (II)}} \,\,\, K=-\frac{a_3}{4}+\frac{a_0}{( \sqrt{q^1}+\sqrt{q^2})^3}.
\end{equation}

The corresponding quantum integral $\hat{F}$ can be chosen in the
form (\ref{q_ops}) with $v^1=q^2, \, v^2=q^1$ and
\begin{eqnarray}
\mbox{\textrm{(I)}} && \phi^1=k\sqrt{-\frac{f(q^1) f(q^2)}{(q^1-q^2)^2}}\, , \ \ \ \  \phi^2=-k \sqrt{-\frac{f(q^1) f(q^2)}{(q^1-q^2)^2}}\, ,    \nonumber\\[3mm]
&&  \varphi=\mu q^1 q^2 -k B (q^1+q^2);\\
\mbox{\textrm{(II)}} &&  \phi^1=k \frac{ \sqrt{-f(q^1) f(q^2)}}{\sqrt{q^1 q^2}-q^2}\, , \ \ \ \  \phi^2=k \frac{ \sqrt{-f(q^1) f(q^2)}}{\sqrt{q^1 q^2}-q^1}\,  ,  \nonumber\\[2mm]
&& \varphi= -\frac{\mu\sqrt{q^1 q^2}}{\sqrt{q^1}+\sqrt{q^2}}-kB\left( \sqrt{q^1}+\sqrt{q^2} \right)^2,
\end{eqnarray}
where $k=-4B/a_3.$
\end{Theorem}

The proof is rather lengthy and technical. We present it now with all the details, mainly following the unpublished work  \cite{FSV}.

\section{Proof of the local classification}

Since the classical and quantum integrability conditions coincide in our case, we will consider for simplicity the classical case,
assuming that the Hamiltonian $H$ and integral $F$ are reduced to the diagonal form
$$
   H=g^{11} \left(p_1-A_1\right)^2+ g^{22}
\left(p_2-A_2\right)^2+h,
$$
$$
   F= g^{11} v^1\left(p_1-A_1\right)^2+ g^{22}v^2
\left(p_2-A_2\right)^2 +\phi^1 \left(p_1-A_1\right)
+\phi^2\left(p_2-A_2\right) + \varphi,
$$ 
where all the coefficients are functions of the local coordinates $(q^1,q^2)$. 

We assume also that the magnetic density $$
B=\sqrt{g^{11} g^{22}} \left( \partial_1 A_2 -\partial_2 A_1\right), \label{mag}
$$
is a non-zero constant. As we have seen in that case the classical and quantum integrability conditions coincide.


Without loss of generality locally we can take $v^1=q^2$, $v^2=q^1$. By integrability condition (C2), we must have metric of St\"ackel form
\begin{equation}
\label{staekel}
ds^2=\frac{q^1-q^2}{f_1(q^1)}(dq^1)^2+\frac{q^2-q^1}{f_2(q^2)}(dq^2)^2.
\end{equation}
 In order to make the metric positive definite, we require that $f_1(q^1)$ and $f_2(q^2)$ have different sign. Now, we use condition (C5)
$$ \partial_1 \varphi - v^1 \partial_1 h - \frac{\phi^2}{\sqrt{g^{11} g^{22}}} B =0,$$
$$ \partial_2 \varphi - v^2 \partial_2  h + \frac{\phi^1}{\sqrt{g^{11} g^{22}}} B =0.$$
The consistency condition gives
\begin{equation}
\phi^1 \partial_1 B + \phi^2 \partial_2 B+ \sqrt{g^{11} g^{22}} \left(v^2-v^1\right)\left( \frac{\partial_2 g^{11}}{g^{11}} \partial_1 h +\frac{\partial_1 g^{22}}{g^{22}}\partial_2 h-\partial_1 \partial_2 h \right)  =0. \label{consistency}
\end{equation}

\begin{remark} 
{\it Note that this condition coincides with the quantum integrability condition $(C6)^*$ given by (\ref{qcond}) with the
roles of $h$ and $B$ interchanged.  Thus we see an interesting {\it duality} between the potential $h$ and the magnetic field
density $B$, which holds {\it only in the quantum case}.

It is interesting that the self-duality conditions $B=\pm h$ appear as the factorisability condition for the Hamiltonian in the work by Ferapontov and Veselov \cite{Fer-Ves}.}
\end{remark}

Since we assumed that $B$ is constant, this relation reduces to
$$
\left(q^1-q^2\right) \partial_1 \partial_2 h-\partial_1 h+\partial_2 h=0,
$$
which can be simplified to
\begin{equation}
\partial_1 \partial_2 \left[ \left(q^1-q^2 \right) h \right]=0.\label{h-pde}
\end{equation}\\
Now assume that $h$ is not a constant.
Solving (\ref{h-pde}) we get
\begin{equation}
h=\frac{a(q^1)-b(q^2)}{q^1-q^2}, \label{h-solution}
\end{equation}
where $a$ and $b$ are arbitrary functions, and
\begin{equation}
\phi^j=-\frac{\partial_i h}{\partial_j h} \phi^i,\ \ \ \ \ \ i\neq j,\ \ \ \ \ i,j=1,2.\label{phi-relation}
\end{equation}
From condition (C3):
$$
\partial_1 \phi^1 = \frac{1}{2 g^{11}} \left( \phi^1 \partial_1 g^{11} + \phi^2 \partial_2 g^{11}\right), \ \ \ \ \  \partial_2 \phi^2 = \frac{1}{2 g^{22}} \left( \phi^1 \partial_1
g^{22} + \phi^2 \partial_2 g^{22}\right),
$$
after rearranging terms and using relation (\ref{phi-relation}), we have
$$
\phi^1=\exp \left[ \int\frac{1}{2 g^{11}}\left( \partial_1g^{11}-\frac{\partial_1 h \ \partial_2 g^{11}}{\partial_2 h} \right) dq^1 \right],
$$
$$
\phi^2=\exp \left[ \int\frac{1}{2 g^{22}}\left( \partial_2g^{22}-\frac{\partial_2 h \ \partial_1 g^{22}}{\partial_1 h} \right) dq^2 \right].
$$
Using the St\"ackel form of the metric (\ref{staekel}) we deduce that
\begin{equation}
\label{ph1}
\phi^1=\beta(q^2) \sqrt{ \frac{f_1(q^1)}{a(q^1)-b(q^2)-(q^1-q^2) b^\prime(q^2) }}
\end{equation}
\begin{equation}
\label{ph2}
\phi^2=\alpha(q^1) \sqrt{ \frac{f_2(q^2)}{b(q^2)-a(q^1)+(q^1-q^2) a^\prime(q^1) }}
\end{equation}
for some arbitrary functions $\alpha$ and $\beta$.
Substituting this back to condition (C6), we have
$$
- \frac{\alpha(q^1)}{\beta(q^2)}= \sqrt{\frac{f_1(q^1) \left( -a(q^1)+b(q^2)+(q^1-q^2) a^\prime(q^1) \right)^3}{f_2(q^2) \left( a(q^1)-b(q^2)-(q^1-q^2) b^\prime(q^2) \right)^3}},
$$
which after taking logarithm and differentiating by $q^1$ and $q^2$, gives
\begin{equation}
\label{ab}
a^{\prime \prime}(q^1)(a(q^1)-b(q^2)-(q^1-q^2) b^\prime(q^2))^3 = b^{\prime \prime}(q^2)
(b(q^2)-a(q^1)+(q^1-q^2) a^\prime(q^1))^3,
\end{equation}
and thus
\begin{equation*}
- \frac{\alpha(q^1)}{\beta(q^2)}= \sqrt{\frac{f_1(q^1) a^{\prime\prime}(q^1)}{f_2(q^2) b^{\prime\prime}(q^2)}}.
\end{equation*}
Rearranging terms and separating $q^1$ and $q^2$, we arrive at the final relation for $\alpha$ and $\beta$
\begin{equation}
-\frac{ \alpha(q^1)}{\sqrt{f_1(q^1) a^{\prime\prime}(q^1)}}=\frac{\beta(q^2)}{\sqrt{f_2(q^2) b^{\prime\prime}(q^2)}}=constant.\label{alpha-beta relation}
\end{equation}
Substituting $q^1=q^2=q$, we have
\begin{eqnarray}
\big(a^{\prime \prime}(q)+b^{\prime \prime}(q) \big) \big( a(q)-b(q) \big)^3 =  0.
\end{eqnarray}
Hence we have the following two cases:
\begin{enumerate}
	\item[A.] $a(q)= b(q)$
	\item[B.] $a^{\prime \prime}(q)=-b^{\prime \prime}(q)$
\end{enumerate}

Case A: \,\,$a(q)= b(q)$

Denote $f(q):=a(q)= b(q)$ and substitute this into equation (\ref{ab}) to have
$$
f^{\prime \prime}(q^1) \left(\frac{ f(q^1)-f(q^2)}{q^1-q^2}-f^\prime(q^2) \right)^3 = f^{\prime \prime}(q^2) \left( - \frac{f(q^1)-f(q^2)}{q^1-q^2}+ f^\prime(q^1) \right)^3.
$$
Then we fix $q^2$ and assume $q^1$ is near to $q^2$. Using Taylor expansion up to $5^{th}$ order derivatives of $f$, we have
\begin{eqnarray}
\frac{ f(q^1)-f(q^2)}{q^1-q^2}&=&f^\prime(q^2) +\frac12 f^{\prime \prime}(q^2) (q^1-q^2)+\frac16 f^{\prime \prime \prime}(q^2) (q^1-q^2)^2  \nonumber \\
&&+\frac{1}{24} f^{(4)}(q^2) (q^1-q^2)^3+\frac{1}{120} f^{(5)}(q^2) (q^1-q^2)^4+ \cdots , \nonumber\\[5mm]
f^\prime(q^1)&=&f^\prime(q^2) + f^{\prime \prime}(q^2) (q^1-q^2)+\frac12 f^{\prime \prime \prime}(q^2) (q^1-q^2)^2  \nonumber \\
&&+\frac{1}{6} f^{(4)}(q^2) (q^1-q^2)^3 +\frac{1}{24} f^{(5)}(q^2) (q^1-q^2)^4+ \cdots , \nonumber\\[5mm]
f^{\prime\prime}(q^1)&=& f^{\prime \prime}(q^2)+ f^{\prime \prime \prime}(q^2) (q^1-q^2)+\frac{1}{2} f^{(4)}(q^2) (q^1-q^2)^2 \nonumber\\
&&+\frac{1}{6} f^{(5)}(q^2) (q^1-q^2)^3+ \cdots . \nonumber
\end{eqnarray}
After the substitution the first coefficients are cancelled, while the cancellation of $(q^1-q^2)^6$ term gives the following necessary condition for $f$: 
\begin{equation}
f^{\prime \prime}(q) \Big( 40  f^{\prime \prime \prime}(q)^3-45 f^{\prime  \prime}(q) f^{\prime \prime \prime}(q) f^{(4)}(q)+9 f^{\prime  \prime}(q)^2 f^{(5)}(q) \Big)=0. \label{f-caseA}
\end{equation}
First we notice that $f^{\prime \prime}(q)$ can not be zero since this will give to a constant potential $h,$ which contradicts our assumption. 

This means that $g(q):=f^{\prime \prime}(q)$ satisfies the equation
\begin{equation}
\frac{40}{9} (g^\prime)^3-5 g\, g^\prime g^{ \prime \prime}+ g^2 g^{ \prime \prime \prime}=0. \label{g-caseA}
\end{equation}
Remarkably this happens to be $n=-\frac23$ case of the following solvable equation:
\begin{equation}
(n-1) (n-2) (y^\prime)^3 +3(n-1) y\, y^\prime y^{ \prime \prime}+y^2 y^{ \prime \prime \prime} =0 \label{PZ}
\end{equation}
with the general solution of the form
$$
[y(x)]^n=c_0+c_1x+c_2 x^2
$$
(see Equation 27 in \cite{PZ}, Section 3.5.3.).

Thus we have
\begin{equation}
f^{\prime \prime}(q)= g(q) = (c_0+c_1 q+c_2 q^2)^{-\frac32}. \label{fprimeprime} \nonumber
\end{equation}
Integrating this twice, we arrive at the following general formula for $f$:
\begin{equation}
f(q)=\frac{4}{4 c_0 c_2 -c_1^2} \sqrt{c_0+c_1 q+c_2 q^2}+C_1 q+C_0, \nonumber
\end{equation}
where $C_0$ and $C_1$ are constants and we assumed that  $4 c_0 c_2 -c_1^2\neq0$. Ignoring the linear term, which only gives a constant shift of the potential, and relabelling the constants we have
\begin{equation}
a(q)=b(q)= \sqrt{c_0+c_1 q+c_2 q^2}.  \nonumber
\end{equation}
In the case, when $4 c_0 c_2 -c_1^2=0$, modulo linear terms we have two subcases:
\begin{equation}
a(q)=b(q)= c q^2, \nonumber
\end{equation}
and 
\begin{equation}
a(q)=b(q)= \frac{c}{q+d}, \nonumber
\end{equation}
where $c$ and $d$ are some constants.

\medskip
Case B: $a^{\prime \prime}(q)=-b^{\prime \prime}(q)$

In that case we have that
$$
a(q)=-b(q)+C_1 q+C_0.
$$
Denote $f(q):=b(q), \, g(q):=2f(q)-C_1 q-C_0$ then, similarly to the previous case, Taylor expansion in the equation (\ref{ab}) 
leads to the following differential equation for $g$:
\begin{equation}
3g^2 g^\prime g^{\prime\prime}+g^3 g^{\prime\prime\prime}=0.\label{g-relation1}
\end{equation}
Trivial solution $g\equiv0$  leads to the constant potential, so we can divide equation (\ref{g-relation1}) by $g^2$ to get
$$
3g^\prime g^{\prime\prime}+g g^{\prime\prime\prime}=0,
$$
which has the general solution
\begin{equation}
g(q)= \sqrt{c_0+c_1 q+c_2 q^2}, \nonumber
\end{equation}
where $c_0$, $c_1$, $c_2$ are arbitrary constants. Hence modulo linear terms
$$
a(q)=- \frac12 \sqrt{c_0+c_1 q+c_2 q^2}=-b(q).
$$
One can check that this case does not lead to any new solutions compared to case A.

Thus we have the following three different cases to analyse:
	\begin{enumerate}
		\item[(1)] $ a(q)=b(q)= \sqrt{c_0+c_1 q+c_2 q^2}$, \vspace{3mm}
		\item[(2)] $ a(q)=b(q)=c q^2$, \vspace{3mm}
		\item[(3)] $ a(q)=b(q)=\displaystyle \frac{c}{q+d}.$\vspace{2mm}
	\end{enumerate}
\medskip	
Case (1): $\  a(q)=b(q)=\sqrt{c_0+c_1 q+c_2 q^2}$ 

Without loss of generality we can reduce this case to 2 subcases
\begin{enumerate}
	\item[(i)] $a(q)=b(q)=\mu\sqrt{q}$, 
	\item[(ii)] $a(q)=b(q)=\sqrt{c_0+c_2 q^2}$.  
\end{enumerate}
\medskip
Subcase (i): $\ \ a(q)=b(q)=\mu\sqrt{q}$

We have
$$
a(q^1)=\mu\sqrt{q^1},\,  b(q^2)=\mu\sqrt{q^2},\,
h=\frac{\mu\sqrt{q^1}-\mu\sqrt{q^2}}{q^1-q^2}=\frac{\mu}{\sqrt{q^1}+\sqrt{q^2}}.
$$
From equation (\ref{alpha-beta relation})
\begin{equation*}
\alpha (q^1)=-\frac{\tilde{k}}{2}  \sqrt{- (q^1)^{-\frac32} f_1(q^1)}\ , \ \ \ \ \beta(q^2)= \frac{\tilde{k}}{2} \sqrt{- (q^2)^{-\frac32} f_2(q^2)}\ ,
\end{equation*}
with some constant $\tilde{k}$. Moreover, from (\ref{ph1}) and (\ref{ph2}) we have
\begin{equation}
\phi^1=k \frac{ \sqrt{-f_1 f_2}}{\sqrt{q^1 q^2}-q^2}, \ \ \ \  \phi^2=k \frac{ \sqrt{-f_1 f_2}}{\sqrt{q^1 q^2}-q^1}, \quad k=\tilde k/\sqrt{\mu}.\nonumber
\end{equation}
Substituting them into the last unused condition (C4), we have
\begin{eqnarray}
\frac{4 B}{k} (q^1-q^2) \left( \sqrt{q^1}-\sqrt{q^2} \right)^2  &=&-\left( \sqrt{q^1}-\sqrt{q^2} \right) \left( \frac{f_1^\prime(q^1)}{\sqrt{q^1}}+\frac{f_2^\prime(q^2)}{\sqrt{q^2}} \right) \nonumber\\
& -&\!\!\!\! \left( \sqrt{\frac{q^2}{q^1}}-2 \right) \frac{f_1(q^1)}{q^1}+\left( \sqrt{\frac{q^1}{q^2}}-2  \right) \frac{f_2(q^2)}{q^2}. \nonumber
\end{eqnarray}
Changing coordinate $\sqrt{q^1}=x ,\sqrt{q^2}=y$ and $f(z^2)=z^3 F(z)$, we have
\begin{eqnarray}
\ \ \ \frac{8 B}{k} (x+y)(x-y)^3  =- (x-y) \left( x F^\prime(x)+y F^\prime(y) \right)+(x+y) (F(x)-F(y)). \nonumber
\end{eqnarray}
Applying the operator $\frac{\partial^3}{\partial x^2 \partial y}$ to this relation yields
\begin{equation*}
\frac{d}{dx}\left(x^3 F^{\prime \prime} (x)\right) = -\frac{96 B}{k} x^3.
\end{equation*}
Solving this ODE gives
\begin{equation*}
F(x)=-\frac{4 B}{k} x^3+a_2 x+\frac{a_1}{x}+a_0,
\end{equation*}
where $a_i$ are constants. Therefore we can derive the functions $f_1$ and $f_2$ in the metric:
\begin{equation}
\label{f}
f_1(q)=f_2(q)=-\frac{4 B}{k} q^3+a_2 q^2+a_1q+a_0 q^{\frac32}.
\end{equation}
To find the potential in the integral $F$  we use condition (C5):
\begin{eqnarray}
\partial_1 \varphi=v^1 \partial_1 h+\frac{\phi^2}{\sqrt{g^{11} g^{22}}} B=q^2 \partial_1 h -k B\left( 1+\sqrt{\frac{q^2}{q^1}} \right), \nonumber\\
\partial_2 \varphi=v^2 \partial_2 h-\frac{\phi^1}{\sqrt{g^{11} g^{22}}} B=q^1\partial_2 h-k B \left( 1+\sqrt{\frac{q^1}{q^2}} \right), \nonumber
\end{eqnarray}
which now has solution
\begin{equation*}
\varphi(q^1,q^2)= -\frac{\mu\sqrt{q^1 q^2}}{\sqrt{q^1}+\sqrt{q^2}}-kB\left( \sqrt{q^1}+\sqrt{q^2} \right)^2.
\end{equation*}

Subcase (ii): $\ \ a(q)=b(q)=\sqrt{c_0+c_2q^2}$

We assume for simplicity that $c_0=c$, $c_2=1$, so
$$
a(q^1)=\sqrt{c+(q^1)^2}, \, b(q^2)=\sqrt{c+(q^2)^2},\,
h=\frac{\sqrt{c+(q^1)^2}-\sqrt{c+(q^2)^2}}{q^1-q^2}.
$$
Again from (\ref{ph1}) (\ref{ph2}) and (\ref{ab}) we have
\begin{eqnarray}
\phi^1=&&k \sqrt{ \frac{-f_1 f_2}{\left(c+(q^2)^2\right) \left( c+q^1q^2-\sqrt{c+(q^1)^2} \sqrt{c+(q^2)^2} \right)}}\ , \nonumber\\
\phi^2=&-&k \sqrt{ \frac{-f_1 f_2}{\left(c+(q^1)^2\right) \left( c+q^1q^2-\sqrt{c+(q^1)^2} \sqrt{c+(q^2)^2} \right)}}\ ,\nonumber
\end{eqnarray}
Similarly to the previous subcase we have
\begin{eqnarray}
&& \frac{4B}{k} (q^1-q^2) \sqrt{c+q^1q^2-\sqrt{c+(q^1)^2} \sqrt{c+(q^2)^2}}  \nonumber\\
=&&-\frac{f_1^\prime(q^1)}{\sqrt{c+(q^1)^2}}-\frac{f_2^\prime(q^2)}{\sqrt{c+(q^2)^2}}\nonumber\\
+&&\frac {c(q^2-q^1)+3q^1\left( c+q^1q^2-\sqrt{c+(q^1)^2} \sqrt{c+(q^2)^2} \right)}{(c+(q^1)^2)^{\frac32}\left( c+q^1q^2-\sqrt{c+(q^1)^2} \sqrt{c+(q^2)^2} \right)} \,f_1(q^1) \nonumber\\
+&&\frac{c(q^1-q^2)+3q^2\left( c+q^1q^2-\sqrt{c+(q^1)^2} \sqrt{c+(q^2)^2} \right)}{\left(c+(q^2)^2\right)^{\frac32}\left( c+q^1q^2-\sqrt{c+(q^1)^2} \sqrt{c+(q^2)^2} \right)} \,f_2(q^2).\nonumber
\end{eqnarray}
Making the substitution
\begin{equation*}
f_1(q^1)=(c+(q^1)^2) F_1(q^1), \ \ \ f_2(q^2)=(c+(q^2)^2) F_2(q^2),
\end{equation*}
we can simplify above equation to be
\begin{eqnarray}
&& \frac{4B}{k} (q^1-q^2) \sqrt{c+q^1q^2-\sqrt{c+(q^1)^2} \sqrt{c+(q^2)^2}}  \nonumber\\[4mm]
=&&-\sqrt{c+(q^1)^2} \,F_1^\prime(q^1)-\sqrt{c+(q^2)^2} \,F_2^\prime(q^2)\nonumber\\[3mm]
&+&\frac {q^2 \sqrt{c+(q^1)^2}-q^1 \sqrt{c+(q^2)^2}}{ \ c+q^1q^2-\sqrt{c+(q^1)^2} \sqrt{c+(q^2)^2} \  } \,F_1(q^1)\nonumber\\[2mm]
&+&\frac{q^1 \sqrt{c+(q^2)^2}-q^2 \sqrt{c+(q^1)^2}}{ \ c+q^1q^2-\sqrt{c+(q^1)^2} \sqrt{c+(q^2)^2} \ } \,F_2(q^2).\nonumber
\end{eqnarray}
We now differentiate this with respect to $q^1$ and $q^2$ to have
\begin{eqnarray}
-&&\!\!\!\!\!\!\!\!\!\!\!\! \frac{B}{ck}\frac{ (q^1-q^2) \left[3c(q^1-q^2)^2+2(c+3q^1q^2)\left(c+q^1q^2-\sqrt{c+(q^1)^2} \sqrt{c+(q^2)^2}\right)\right]}{\sqrt{c+q^1q^2-\sqrt{c+(q^1)^2} \sqrt{c+(q^2)^2}}}  \nonumber\\
=&&-\sqrt{c+(q^1)^2} \,F_1^\prime(q^1)-\sqrt{c+(q^2)^2} \,F_2^\prime(q^2)\nonumber\\
&+&\frac {q^2 \sqrt{c+(q^1)^2}-q^1 \sqrt{c+(q^2)^2}}{ \ c+q^1q^2-\sqrt{c+(q^1)^2} \sqrt{c+(q^2)^2} \  } \,F_1(q^1)\nonumber\\
&+&\frac{q^1 \sqrt{c+(q^2)^2}-q^2 \sqrt{c+(q^1)^2}}{ \ c+q^1q^2-\sqrt{c+(q^1)^2} \sqrt{c+(q^2)^2} \ } \,F_2(q^2).\nonumber
\end{eqnarray}
Note that these two equations are only different in the left hand side, so we can subtract them to get
\begin{equation}
B\bigg[c(q^1-q^2)^2+2(c+q^1q^2)\left( c+q^1q^2-\sqrt{c+(q^1)^2} \sqrt{c+(q^2)^2}\right)\bigg]=0. \nonumber
\end{equation}
Since we assumed that $B\neq 0,$ we have $$ c(q^1-q^2)^2+2(c+q^1q^2)\left( c+q^1q^2-\sqrt{c+(q^1)^2} \sqrt{c+(q^2)^2} \right)=0,$$ which after simplification reduces to
\begin{equation*}
0=c^2 (q^1-q^2)^4.
\end{equation*}
From this we have $c=0,$ which leads to a constant potential $h$. 

Thus in the subcase (ii) we have no required integrable cases.

\medskip
Case (2): $\ \ a(q)=b(q)=c q^2$ 

Without loss of generality we can assume that $c=1$, so
$$
a(q^1)=(q^1)^2, \,  b(q^2)=(q^2)^2, \, h=q^1+q^2.$$
From equations (\ref{ph1}), (\ref{ph2}) and (\ref{alpha-beta relation}) we have
\begin{equation}
\phi^1=k\sqrt{-\frac{f_1 f_2}{(q^1-q^2)^2}}\, , \ \ \ \  \phi^2=-k \sqrt{-\frac{f_1 f_2}{(q^1-q^2)^2}}, \nonumber
\end{equation}
where $f_1=f_1(q^1), \, f_2=f_2(q^2)$ and $k$ is an arbitrary constant.
 From condition 4), we have
\begin{equation}
\frac{B}{k}=  \frac{f_1(q^1)-f_2(q^2)}{2 (q^1-q^2)^3}-\frac{f_1^\prime(q^1)+f_2^\prime(q^2)}{4 (q^1-q^2)^2}. \label{Gaussian curvature}
\end{equation}
Rearranging and putting $q^1=q^2=q$ implies that $f_1(q)=f_2(q):=f(q)$. Hence equation (\ref{Gaussian curvature}) reduces to
\begin{equation}
\frac{4B}{k} (q^1-q^2)^3=  2 \left( f(q^1)-f(q^2) \right)-(q^1-q^2) \left(f^\prime (q^1)+f^\prime (q^2) \right). \nonumber
\end{equation}
Applying $\frac{\partial^3}{\partial q^{1^2} \partial q^{2}}$ to the above equation gives
$
f^{\prime \prime \prime}=-24B/k,
$
implying that
\begin{equation*}
f_1(q)=f_2(q)=- \frac{4B}{k} q^3+a_2 q^2+a_1 q+a_0.
\end{equation*}
The Gaussian curvature $K$ in this case is a constant equals to $\frac BK.$\\
As before, to find the integral $F$ we use condition 5): 
$$
\partial_1 \varphi=v^1 \partial_1 h+\frac{\phi^2}{\sqrt{g^{11} g^{22}}} B=q^2-k B,\,\,
\partial_2 \varphi=v^2 \partial_2 h-\frac{\phi^1}{\sqrt{g^{11} g^{22}}} B=q^1-k B,
$$
so that
\begin{equation}
\varphi(q^1,q^2)=q^1 q^2 -k B (q^1+q^2). \nonumber
\end{equation}

\medskip
Case (3):  $\ \ a(q)=b(q)=\frac{c}{q+d}$

We can assume for simplicity that $c=1$ and $d=0$, so
$$
a(q)=b(q)=1/q, \,  h=-1/q^1 q^2.
$$
From (\ref{ph1}) and (\ref{ph2}) we have
\begin{equation}
\phi^1=k \sqrt{-\frac{q^1 f_1 f_2}{q^2 (q^1-q^2)^2}}\, , \ \ \ \  \phi^2=-k \sqrt{-\frac{q^2 f_1 f_2}{q^1 (q^1-q^2)^2}}\, . \nonumber
\end{equation}
Applying the operator $\frac{\partial^4}{\partial q^{1^2} \partial q^{2^2}}$ to condition (C4) in this case, we have
\begin{equation*}
\Big(5 (q^1)^3+(q^1)^2 q^2 -q^1 (q^2)^2-5(q^2)^3\Big) B =0,
\end{equation*}
which means that in this case magnetic field is zero.

Thus we have shown that only cases (1)(i) and (2) lead to the integrable systems with non-zero constant magnetic field and non-constant potential. This completes the proof of Theorem 1.

We should emphasize that this is a local classification and all these metrics are incomplete. 
We are going to show now that under certain assumptions on the parameters these systems can be extended to the analytic integrable systems on a topological sphere, thus presenting some integrable generalisations of the Dirac magnetic monopole.

\section{Case I: Dirac magnetic monopole in harmonic field}

To understand the global geometry of the case I we should consider two major subcases, when the cubic polynomial $f(q)=a_3 q^3+a_2 q^2+a_1 q+a_0$ has

I a) three distinct real roots;

II b) one real root and two complex conjugated roots.

It is easy to check that the metric (\ref{metric}) is positive definite and has positive Gaussian curvature $K$ only in the case I a) with $a_3<0.$

Let us show that in this case this metric is simply the standard metric on a round sphere $S^2 \subset \mathbb R^3.$
Without loss of generality we can restrict ourselves to the case $a_3=-4$ corresponding to the unit sphere.

%
Consider a sphere given in Cartesian coordinates $x_1$, $x_2$, $x_3$ in $\mathbb R^3$ by the equation
\begin{equation}
  x_1^{\,2}+x_2^{\,2}+x_3^{\,2}=1, \nonumber
\end{equation}
and introduce, following C. Neumann, the spherical elliptic coordinates as the roots $q^1$, $q^2$ of the quadratic equation
\begin{equation}
  \phi(q)=\frac{x_1^{\,2}}{\alpha_1-q}+\frac{x_2^{\,2}}{\alpha_2-q}+\frac{x_3^{\,2}}{\alpha_3-q}=0, \label{quadforq}
\end{equation}
\\[-4mm]
where $\alpha_1$, $\alpha_2$, $\alpha_3$ are arbitrary constants (see \cite{Moser,N}).  Rewrite the quantity $\phi$ in terms of the roots $q^1$, $q^2$ as follows:\\[-3mm]
\begin{equation}
  \phi(q) =\frac{(q-q^1)(q-q^2)}{(\alpha_1-q)(\alpha_2-q)(\alpha_3-q)} \nonumber
\end{equation}
\\[-3mm]
and computing the residues we come to the following expression of the Cartesian coordinates $x_1$, $x_2$, $x_3$ and the spherical elliptic coordinates $q^1$, $q^2$:\\[-4mm]
\begin{eqnarray}
  \ \ x_1^{\,2}= \frac{(\alpha_1-q^1)(\alpha_1-q^2)}{(\alpha_1-\alpha_2)(\alpha_1-\alpha_3)}, \,
  x_2^{\,2}= \frac{(\alpha_2-q^1)(\alpha_2-q^2)}{(\alpha_2-\alpha_1)(\alpha_2-\alpha_3)}, \,
  x_3^{\,2}= \frac{(\alpha_3-q^1)(\alpha_3-q^2)}{(\alpha_3-\alpha_1)(\alpha_3-\alpha_2)}. \nonumber
\end{eqnarray}
A simple calculation shows then that in the elliptic coordinates $q^1$, $q^2$ the metric takes the form
\begin{equation}
  ds^2=  \frac{q^1-q^2}{4(\alpha_1-q^1)(\alpha_2-q^1)(\alpha_3-q^1)} (dq^1)^2+  \frac{q^2-q^1}{4(\alpha_1-q^2)(\alpha_2-q^2)(\alpha_3-q^2)} (dq^2)^2, \nonumber
\end{equation}
which is of St\"ackel type (\ref{metric}) with cubic polynomial
\begin{eqnarray}
  f(x)= 4(\alpha_1-x)(\alpha_2-x)(\alpha_3-x) \nonumber
\end{eqnarray}\\[-6mm]
having 3 real roots. 

Note that if we order the roots and the elliptic coordinates by $$\alpha_1>q^1>\alpha_2>q^2>\alpha_3,$$ then we have general case of metrics in class I a) with $x=q^1, \, y=q^2.$
The degenerate case, when two of the roots of cubic $f$ collide, corresponds to the usual spherical coordinates on sphere.

Let us show now that in terms of Cartesian coordinates the potential $h=\mu(q^1+q^2)$ is quadratic. We have by definition
$$
q^2-\Big[ (\alpha_2+\alpha_3) x_1^{\,2}+(\alpha_1+\alpha_3) x_2^{\,2}+(\alpha_1+\alpha_2) x_3^{\,2} \Big]q
 +(\alpha_2 \alpha_3 x_1^{\,2}+\alpha_1 \alpha_3x_2^{\,2}+\alpha_1 \alpha_2 x_3^{\,2})=0,
 $$
which implies that
$
  q^1+q^2= (\alpha_2+\alpha_3) x_1^{\,2}+(\alpha_1+\alpha_3) x_2^{\,2}+(\alpha_1+\alpha_2) x_3^{\,2}.
$
Thus the potential $h=\mu(q^1+q^2)$ is a quadratic function of $x_1$, $x_2$, $x_3$, which could be chosen arbitrary.

\begin{Theorem} 
Integrable systems of type I a) with $a_3<0$ can be extended to the Dirac magnetic monopoles on the round sphere in the external harmonic field with arbitrary quadratic potential.
They are equivalent to the classical integrable Clebsch systems considered on the co-adjoint orbits of the Euclidean group $E(3).$
\end{Theorem}

Indeed, it is well-known that the Dirac magnetic monopole in the external harmonic field is equivalent to a special Clebsch integrable case of the rigid body motion in the infinite ideal fluid (see \cite{V}).

Recall that the Kirchhoff equations for such a motion are simply Euler equations on the dual space $e(3)^*$ of the Lie algebra of the isometry group $E(3)$ of Euclidean space $\mathbb R^3$ (see e.g. Perelomov \cite{Perelomov}). 
The corresponding variables $M_i, x_i, \, i=1,2,3$ have the canonical Lie-Poisson brackets
\begin{equation}
\label{e3}
\left\{M_i, M_j\right\} = \epsilon_{ijk} M_k, \ \  \left\{M_i, x_j\right\} = \epsilon_{ijk} x_k, \ \  \left\{x_i, x_j \right\}=0.
\end{equation}
We have two Casimir functions 
$$
C_1=|x|^2,\quad C_2=(M,x).
$$
As it was first pointed out by S.P. Novikov and Schmelzer \cite{NS}, the symplectic leaf with $C_1=|x|^2=1, C_2=(M,x)=\nu$ is symplectically isomorphic to the cotangent bundle of the unit sphere $T^*S^2$
with additional Dirac magnetic field with density $B=\nu$. 

In the coordinates $M,x$ the Hamiltonian and the integral of the corresponding Clebsch system have the form
\begin{equation}
\label{clebsch}
H=|M|^2-\mu(\alpha_1 x_1^2+\alpha_2 x_2^2+\alpha_3 x_3^2),
\end{equation}
$$
F=\alpha_1 M_1^2+\alpha_2 M_2^2+\alpha_3 M_3^2+\mu(\alpha_2\alpha_3 x_1^2+\alpha_1\alpha_3 x_2^2+\alpha_1\alpha_2 x_3^2).
$$
To get the quantum version one should simply replace $M,x$ by $\hat M, \hat x$ with the commutation relations
$$
[\hat M_i, \hat M_j] = \epsilon_{ijk} \hat M_k, \ \  [\hat M_i, \hat x_j] = \epsilon_{ijk} \hat x_k, \ \  [\hat x_i, \hat x_j]=0.
$$
Note that there is no ordering problem since both Hamiltonian and integral written only in terms of the squares of variables.

In the remaining cases of type I we have different versions of elliptic coordinates on the hyperbolic plane in external harmonic field, see e.g. \cite{V2}.

Let us consider here only the most degenerate case when $f(x)=4x^3.$
Making change of variables
$X=(q^1)^{-1/2},\,Y=(q^2)^{-1/2},$
we have
$$ds^2=(\frac{1}{X^2}+\frac{1}{Y^2})(dX^2+dY^2).$$
Denote $w=X+iY$ and $z=w^2=X^2-Y^2+2iXY=u+iv$, then
$$ds^2=\frac{X^2+Y^2}{X^2Y^2}(dX^2+dY^2)=\frac{4w\bar{w}}{\Im(w^2)^2}dwd\bar{w}=\frac{dzd\bar{z}}{\Im(z)^2}=\frac{du^2+dv^2}{v^2},$$
which is the canonical hyperbolic metric on the upper half plane.
The potential $h$ in $u,v$-coordinates is
$$
h=\mu(q^1+q^2)=\mu\frac{Y^2-X^2}{X^2Y^2}=-\frac{4\mu u}{v^2}.
$$

\section{Case II: new integrable generalisations of Dirac monopole}

Let us first of all rewrite the formulas in more convenient variables 
$$
x_1=\sqrt{q^1}, \, x_2=\sqrt{q^2}.
$$
Then metric (\ref{metric}) takes the form
\begin{equation}
\label{metric1}
ds^2=4\frac{x_1^2-x_2^2}{P(x_1)}dx_1^2+4\frac{x_2^2-x_1^2}{P(x_2)}dx_2^2
\end{equation}
with 
\begin{equation}
\label{P}
P(x)=a_3x^4+a_2x^2+a_0x+a_1
\end{equation}
(note an unusual order of the coefficients). 
The Gaussian curvature in the new coordinates is 
\begin{equation}
\label{K}K=-\frac{a_3}{4}+\frac{a_0}{(x_1+x_2)^3}.
\end{equation}
The electric potential $h$ becomes
\begin{equation}
\label{h}
h=\frac{\mu}{x_1+x_2},
\end{equation}
while the magnetic potential $A$ is determined by
\begin{equation}
\label{A}
\partial_1 A_2-\partial_2 A_1=4B\frac{x_1^2-x_2^2}{\sqrt{-P(x_1)P(x_2)}}.
\end{equation}
The integral $F$ has the form (\ref{q_ops}) with
\begin{equation}
\label{phi}
\phi^1=k \frac{\sqrt{-P(x_1)P(x_2)}}{2(x_1-x_2)}=-\phi^2, \quad \varphi= -\frac{\mu x_1x_2}{x_1+x_2}-kB(x_1+x_2)^2,
\end{equation}
where as before $k=-4B/a_3.$

To study the regularity condition we can assume without loss of generality that $a_3<0$ and $a_0\leq 0$.
For the analysis of the special case $a_0=0$ we refer to our paper \cite{VY}, so let us assume now that $a_0<0.$

One can show that in order to define regular system on a sphere the polynomial $P(x)$ must have 4 real roots, which we denote $\beta_i, \, i =1,2,3,4:$
$$
P(x)=a_3x^4+a_2x^2+a_0x+a_1=a_3(x-\beta_1)(x-\beta_2)(x-\beta_3)(x-\beta_4).
$$
We assume also that there are no multiple roots and that 
$\beta_1>\beta_2>\beta_3>\beta_4,$ such that
$$
\beta_1+\beta_2+\beta_3+\beta_4=0.
$$
Simple arguments show that we have that actually $\beta_1>\beta_2>0>\beta_3>\beta_4$ and that
\begin{equation}
\label{roots}
\beta_1+\beta_4<0, \quad \beta_2+\beta_3>0
\end{equation} 
(see Figure 1).

\begin{figure}[h]
		\centering	
		\includegraphics[width=2in,height=1.85in]{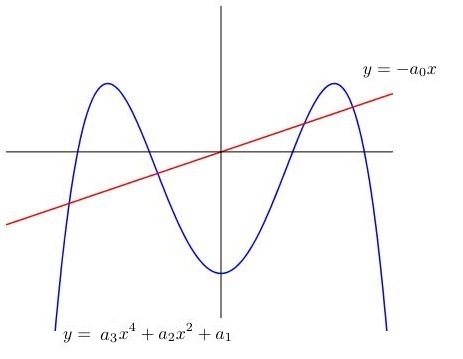} \quad \includegraphics[width=2in,height=1.85in]{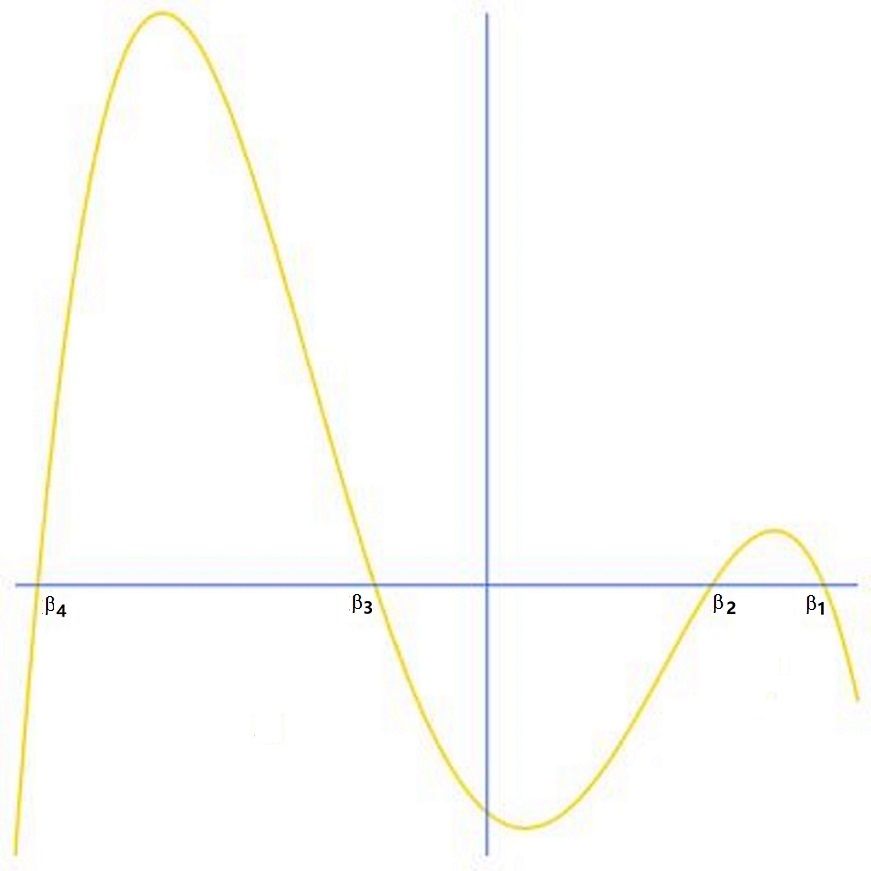}
		\caption{Graph and zeroes of $P(x)$}
		\label{Case-II}
	\end{figure}

The algebraic conditions on the coefficients of the quartic polynomial (\ref{P}) for having 4 distinct real roots are
$$
\Delta>0, \,\, a_2 a_3<0, \,\, 4a_1a_3-a_2<0,
$$
where $\Delta$ is the discriminant of $P(x)=0$:
$$
\Delta=256a_1^3a_3^3-128a_1^2a_2^2a_3^2+144a_0^2a_1a_2a_3^2-27a_0^4a_3^2+16a_1a_2^4a_3-4a_0^2a_2^3a_3,
$$
or, under our assumption that $a_3<0$,
\begin{equation}
\label{condroots1}
a_2>0, \,\, a_3<0, \,\, a_1<\frac{a_2}{4a_3}<0,
\end{equation} 
\begin{equation}
\label{condroots2}
256a_1^3a_3^2-128a_1^2a_2^2a_3+144a_0^2a_1a_2a_3-27a_0^4a_3+16a_1a_2^4-4a_0^2a_2^3<0.
\end{equation} 

Under these assumptions we can make change of variables
\begin{equation}
\label{u}
u_1=\int_{\beta_2}^{x_1}\frac{2dx}{\sqrt{P(x)}},\,\,\,\,\, u_2=\int_{\beta_2}^{x_2} \frac{2dx}{\sqrt{-P(x)}}
\end{equation} 
with $x_1\in[\beta_2,\beta_1], \, x_2 \in[\beta_3,\beta_2].$

We can express the variables $x_1,x_2$ via $u_1,u_2$ using the elliptic function $\mathcal Q(z)$ defined as the inversion $w=\mathcal Q(z)$ of the elliptic integral
\begin{equation}
\label{Q}
z=\int_{\beta_2}^{w}\frac{2d\xi}{\sqrt{P(\xi)}}=\int_{\beta_2}^{w}\frac{2d\xi}{\sqrt{a_3\xi^4+a_2\xi^2+a_0\xi+a_1}},
\end{equation} 
 as follows
 \begin{equation}
\label{u2}
x_1=Q_1(u_1):=\mathcal Q(u_1),\,\,\,\,\, x_2=Q_2(u_2):=\mathcal Q(iu_2).
\end{equation} 
The elliptic function $\mathcal Q(z)$ is even, of order 2 and 
has two periods: real $2K_1$ and pure imaginary $2iK_2$, where
 \begin{equation}
\label{periods}
K_1=\int_{\beta_2}^{\beta_1}\frac{2d\xi}{\sqrt{P(\xi)}}, \,\,\, K_2=\int_{\beta_2}^{\beta_3}\frac{2d\xi}{\sqrt{-P(\xi)}}.
\end{equation} 
It satisfies the differential equation
$$
4\mathcal Q'^2=P(\mathcal Q)=a_3\mathcal Q^4+a_2\mathcal Q^2+a_0\mathcal Q+a_1
$$
and can be expressed via the standard Weierstrass elliptic function $\wp(z)$.

In particular, when $a_0=0$ we have
$$
4\mathcal Q'^2=a_3(\mathcal Q^2-\beta_1^2)(\mathcal Q^2-\beta_2^2)
$$
and $\mathcal Q$ can be written as one of the Jacobi's elliptic functions \cite{WW}:
$$
\mathcal Q=\beta_2 \, \textit{sn}(\alpha (z-\beta_2);k),\,\, \alpha=\sqrt{a_3}\beta_1/2, \,\, k=\beta_2/\beta_1.
$$
 In the new coordinates the metric (\ref{metric1}) takes the form
\begin{equation}
\label{metric2}
ds^2=(Q_1^2(u_1)-Q_2^2(u_2))(du_1^2+du_2^2),
\end{equation} 
and the potential is
\begin{equation}
\label{pot2}
h=\frac{\mu}{Q_1(u_1)+Q_2(u_2)}.
\end{equation} 

Consider now the real torus 
$$\mathbb T^2=\mathbb R^2(u_1,u_2)/4K_1 \mathbb Z\oplus 4K_2 \mathbb Z,$$
identifying the points $(u_1,u_2)$ and  $(u_1+4K_1m,u_2+4K_2n), \,\, m,n \in \mathbb Z.$

Formula (\ref{metric2}) defines a semi-positive metric on $\mathbb T^2$. Indeed, $$Q_1^2(u_1)-Q_2^2(u_2)=x_1^2-x_2^2=(x_1+x_2)(x_1-x_2)\geq 0,$$
since $x_1+x_2\geq \beta_2+\beta_3>0$ by (\ref{roots}) and $x_1\geq x_2.$ The potential $h$ is regular everywhere on the torus, since the denominator $Q_1(u_1)+Q_2(u_2)=x_1+x_2$ is always positive.

Thus (\ref{metric2}) fails to be a Riemannian metric on $\mathbb T^2$ only at the points when $x_1=x_2=\beta_2,$ which correspond to $(u_1,u_2)=(0,0)$ and three half-periods $(2K_1,0), (0,2K_2), (2K_1,2K_2)$ of the torus.

Note that the functions $Q_1$ and $Q_2$ are even, so the metric and the potential are invariant under the involution
$$
\sigma:(u_1,u_2)\rightarrow (-u_1,-u_2),
$$
having exactly those 4 points fixed. The quotient $\mathbb T^2/\sigma=\mathbb S^2$ is a topological sphere (see Fig. 2, where we are using octahedron to represent it).

\begin{figure}[h]
		\centering	
		\includegraphics[width=5.6in,height=1.85in]{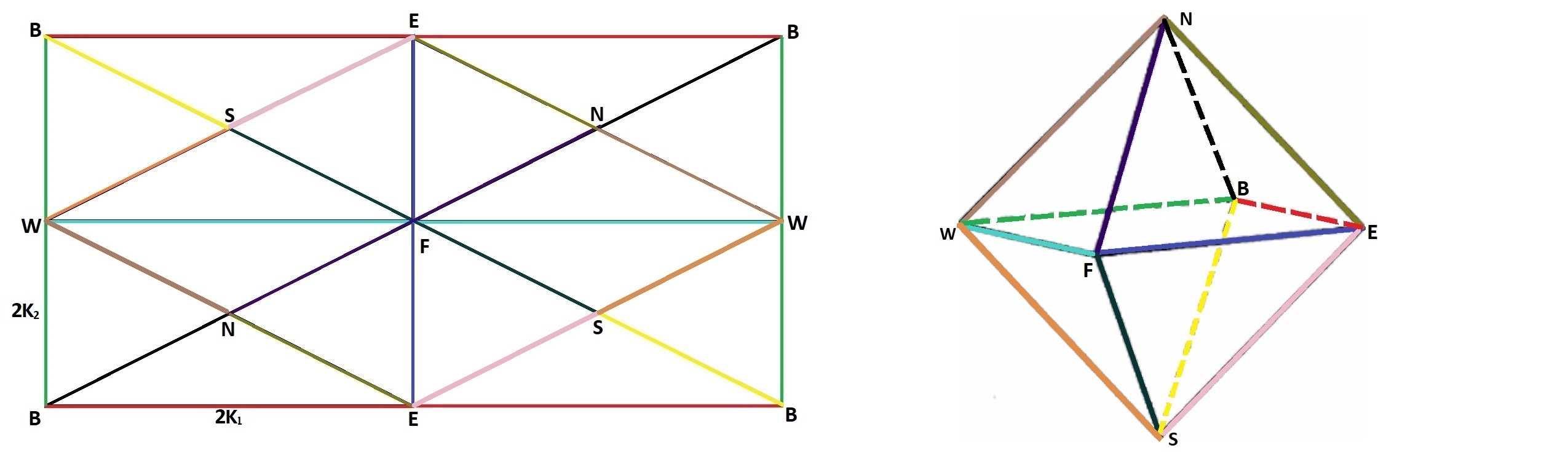}
		\caption{Octahedron as a quotient of torus by involution}
		\label{topsphere}
	\end{figure}
	
We claim that the projection $p: \mathbb T^2 \rightarrow \mathbb S^2$ maps the semi-positive metric (\ref{metric2}) to a proper Riemannian metric on  $\mathbb S^2$ with induced smooth structure. Indeed, we need to check only that this works in the vicinity of the 4 fixed points.

Let us check this at the point $(0,0).$ If $x\approx \beta_2$ then $P(x)\approx c(x-\beta_2),\, c=P'(\beta_2),$ 
$$
u_1=\int_{\beta_2}^{x_1}\frac{2dx}{\sqrt{P(x)}}\approx \int_{\beta_2}^{x_1}\frac{2dx}{\sqrt{c(x-\beta_1)}}=\frac{4}{\sqrt{c}}\sqrt{x_1-\beta_2}.
$$
Thus near $(0,0)$ we have 
$
x_1\approx \beta_2+Cu_1^2, \,\, x_2\approx \beta_2-Cu_2^2, \,\, C=\sqrt{c}/4,
$
and thus
$
x_1+x_2\approx 2\beta_2, \,\, x_1-x_2\approx C(u_1^2+u_2^2),\,\, x_1^2-x_2^2 \approx 2C\beta_2(u_1^2+u_2^2).
$

Thus locally metric (\ref{metric2}) has the form
$
ds^2\approx 2C\beta_2(u_1^2+u_2^2)(du_1^2+du_2^2)=2C\beta_2z\bar z dzd\bar z,
$
where we introduced complex coordinate $z=u_1+iu_2.$ The involution $\sigma$ acts by $z\to -z$, so the complex coordinate on the quotient is $w=z^2=v_1+iv_2$, in which metric takes regular form
$
ds^2\approx \frac{1}{2}C\beta_2dwd\bar w=\frac{1}{2}C\beta_2(dv_1^2+dv_2^2).
$

The situation near 3 other fixed points is similar. Thus we have proved
\begin{Theorem} 
Local integrable systems of type II given by (\ref{II}) with parameters, satisfying the conditions (\ref{condroots1}),(\ref{condroots2}), can be extended to smooth generalisations of Dirac magnetic monopole (\ref{H1}),(\ref{F1}),(\ref{A1}) on topological sphere $\mathbb S^2$ with special metric given in terms of elliptic functions by (\ref{metric2}), (\ref{pot2}).
\end{Theorem}

In the quantum case we should add the usual quantisation conditions for the total magnetic flux
\begin{equation}
\label{Quant}
\frac{1}{2\pi}B \int_{\mathbb S^2} d\sigma \in \mathbb Z,
\end{equation}
where $d\sigma$ is the area form on sphere with metric (\ref{metric2}). Geometrically this is the integrality of the first Chern class of the corresponding line bundle \cite{WY}.

In the limiting case $a_0=0$ the metric on the sphere becomes standard, but the potential becomes singular at two points. The corresponding system can be viewed as a new integrable version of Euler two-centre problem and was studied in \cite{VY}.

\begin{Theorem} \cite{VY} 
The system of type II given by (\ref{II}) with $a_0=0$ can be written, similarly to type I, on the dual Lie algebra $e(3)^*$, where the Hamiltonian and integral have the following form 
\begin{equation}
\label{H}
H=\frac{1}{2}|M|^2-\mu\frac{|q|}{\sqrt{R(q)}},
\end{equation}
\begin{equation}
\label{F}
F=A M_1^2+B M_2^2+\frac{2 \sqrt{A B}}{|q|}(M,q) M_3-2\mu\sqrt{AB}\frac{q_3}{\sqrt{R(q)}},
\end{equation}
where
$
R(q)=Aq_2^2+Bq_1^2+(A+B)q_3^2-2\sqrt{AB}|q|q_3
$
and $\mu, A, B$ are parameters satisfying $A>B>0$. 

The corresponding electric potential has two Coulomb-like singularities, so this system can be considered as new integrable two-centre problem on the sphere in the external Dirac magnetic field.
\end{Theorem}


Let us consider now another limiting case when $\beta_1=\beta_2$, assuming for simplicity that $a_3=-1.$ The function $\mathcal Q(u)$ satisfies the equation
$$
4\mathcal Q'^2=-(\mathcal Q-\beta_1)^2 R(\mathcal Q),\quad R(\xi)=(\xi-\beta_3)(\xi-\beta_4).
$$
Solving this equation and putting $u=iu_2$, we have
\begin{equation}
\label{limit2}
x_2=Q_2(u_2)=\beta_1-\frac{4ce^{\frac{1}{2}\sqrt{c}u_2}}{(b+e^{\frac{1}{2}\sqrt{c}u_2})^2-4c},
\end{equation}
where $$b=2\beta_1-\beta_3-\beta_4=4\beta_1>0, \,\, c=R(\beta_1)=(\beta_1-\beta_3)(\beta_1-\beta_4)>0.$$
Note that since  $b^2-4c>0$
 the denominator in (\ref{limit2}) is always positive, $\beta_3\leq Q_2(u_2) < \beta_1=\beta_2$ and $Q_2(u_2)\to \beta_1$ as $u_2 \to \pm \infty.$
 Since
 $$
 \beta_1-\frac{4ce^{\frac{1}{2}\sqrt{c}u}}{(b+e^{\frac{1}{2}\sqrt{c}u})^2-4c}=\beta_1-\frac{4c}{\sqrt{D}\cosh\frac12\sqrt{c}(u-\delta)+2b},
 $$
 where $D=b^2-4c, \delta=\frac{\ln D}{\sqrt{c}},$ we see that $Q_2$ has the symmetry
 $$
 Q_2(2\delta-u)=Q_2(u).
 $$
 
 We have a problem with the first coordinate $x_1$ though, since the second solution of the equation is $Q_1(u_1)\equiv \beta_1$.  
 
 To deal with this issue we consider the limit $\beta_2 \to \beta_1$ more carefully. Namely, let us introduce
 $\varepsilon=\frac12 (\beta_1-\beta_2), \bar \beta=\frac12 (\beta_1+\beta_2),$ so that
 $$-(x-\beta_1)(x-\beta_2)=\varepsilon^2-(x-\bar \beta)^2.$$
 Define now coordinate $u_1$ as the integral
 $$u_1=\int_{\bar\beta}^{x_1} \frac{dx}{\sqrt{\varepsilon^2-(x-\bar \beta)^2}}=\arcsin \frac{x_1-\bar\beta}{\varepsilon},$$
so that the inversion gives
\begin{equation}
\label{limit1}
x_1=\bar\beta+\varepsilon \sin u_1.
\end{equation}
Since we have 
$$
du_1^2=\frac{dx_1^2}{\varepsilon^2-(x_1-\bar \beta)^2}
$$
we see that in coordinates $u_1,u_2$  when $\varepsilon \to 0$ the metric (\ref{metric1})  has the following limit on the cylinder $0\leq u_1 \leq 2\pi, \, u_2 \in \mathbb R$ : 
\begin{equation}
\label{metric3}
ds^2=\frac{4(\beta_1^2-Q_2^2(u_2))}{c}\left [du_1^2+\frac{c}{4}du_2^2\right ].
\end{equation}
where $c=R(\beta_1)=(\beta_1-\beta_3)(\beta_1-\beta_4).$

We claim that this metric can be extended to the sphere. To show consider first the central projection $p$ of the cylinder $x^2+y^2=1$ to the unit sphere $S^2$ given by 
$x^2+y^2+z^2=1.$ 

Parametrising the cylinder as $x=\cos v_1, y=\sin v_1, z=\sinh v_2$ after a simple calculation we have the following form of the metric on the cylinder, induced from the standard metric on the $S^2$:
\begin{equation}
\label{metric4}
ds^2=\frac{1}{\cosh^2 v_2}\left [dv_1^2+ dv_2^2\right].
\end{equation}

Now let us change variables in (\ref{metric3}) as follows
\begin{equation}
\label{change2}
u_1=2\tilde{u}_1, \, u_2=\frac{4}{\sqrt{c}}\tilde{u}_2+\delta,
\end{equation}
so that the metric takes the form
\begin{equation}
\label{metric5}
ds^2=\frac{16(\beta_1^2-\tilde{Q}_2^2(\tilde{u}_2))}{c}\left [d\tilde{u}_1^2+d\tilde{u}_2^2\right ]
\end{equation}
with 
$$
\tilde{Q}_2(\tilde{u})=\beta_1-\frac{4c}{\sqrt{D}\cosh 2\tilde u+2b}.
$$

Since $\beta_1^2-\tilde{Q}_2^2(\tilde{u}_2)$ decays as $A e^{-2 \tilde{u}_2}, \, A=\frac{8\beta_1 c}{\sqrt{D}}$ when $\tilde{u}_2 \to \infty$ (and as $A e^{2\tilde{u}_2}$ when $\tilde{u}_2 \to -\infty$), we see that the asymptotic behaviour of the metric (\ref{metric5}) is the same as the standard metric on the unit sphere (in cylindrical version (\ref{metric4})).

Note that the change $u_1=2\tilde{u}_1$ corresponds to the double covering of the sphere by the cylinder (which is the degeneration of the torus).

The electric potential $h$ in the coordinates $u_1,u_2$ has the form
$$
h=\frac{\mu}{\beta_1+Q_2(u_2)},
$$
while the magnetic potential satisfies
$$
\partial_2 A_1-\partial_1 A_2=B\frac{\beta_1^2-Q_2^2(u_2)}{R(\beta_1)}.
$$
Note that since the right-hand side is independent on $u_1$, we can choose $A_2\equiv 0$ and
$$
A_1=\int_{-\infty}^{u_2} B\frac{\beta_1^2-Q_2^2(\xi)}{R(\beta_1)}d\xi.
$$
Since all the coefficients in the Hamiltonian
$$
H=\frac{R(\beta_1)}{4(\beta_1^2-Q_2^2(u_2))}\left [(p_1-A_1(u_2))^2+\frac{4}{R(\beta_1)}p_2^2\right ]+\frac{\mu}{\beta_1+Q_2(u_2)}
$$
do not depend on $u_1,$ the system has an obvious linear integral $F=p_1,$ and thus is not covered by Theorem 1.


\section{Concluding remarks}

There are several natural questions about new integrable case II, which are still to be answered.

We have an interesting metric on topological $\mathbb S^2$ defined by (\ref{metric2}). Can it be induced from the Euclidean metric by a suitable embedding of $\mathbb S^2$ into $\mathbb R^3$?
If yes, is there an explicit realisation of such a surface?

To study the orbits in the classical version of new system and especially the spectrum of the corresponding quantum problem seems to be a very difficult problem.
Part of the reasons is the non-zero magnetic field, which is prevent the standard use of the separation of variables (see although recent interesting progress in this direction in \cite{MagriS,S}). 

A limiting even case with $a_0=0$ would be easier to study since in that case we have the usual Dirac magnetic monopole with additional electric field \cite{VY}.

\section{Acknowledgements}

We are very grateful to Alexey Bolsinov and Jenya Ferapontov for many useful and stimulating discussions.

The work of A.P. Veselov was supported by the Russian Science Foundation grant no. 20-11-20214.

\end{document}